# The Impact of Tamm Plasmons on Photonic Crystals Technology

Simone Normani,[a] Francesco Federico Carboni,[a] Guglielmo Lanzani,[a,c] Francesco Scotognella[c] and Giuseppe Maria Paternò[c]*

a. Center for Nano Science and Technology@PoliMi, Istituto Italiano di Tecnologia, Via Giovanni Pascoli, 70/3, 20133 Milano, Italy

b. Institute for Photonics and Nanotechnologies (IFN), Consiglio Nazionale delle Ricerche (CNR), Piazza L. da Vinci 32, 20133 Milano, Italy

c. Physics Department, Politecnico di Milano, Piazza Leonardo Da Vinci, 32, 20133 Milano, Italy

*Authors to whom correspondence should be addressed: giuseppemaria.paterno@polimi.it

## Abstract

This review describes hybrid photonic-plasmonic structures based on periodic structures that have metallic coatings or inserts which make use of the Tamm plasmon for sensing applications. The term Tamm plasmon refers a particular resonance resulting from the enhancement of a surface plasmon resonance absorption via coupling to a wavelength-matching photonic bandgap provided by a photonic crystal. Tamm plasmon-based devices come in an ample variety of material and geometric combinations, each designed to perform a specific kind of measurement. While the physical effect is quite well documented and understood, its implementation in devices is still a rapidly developing and thriving field, which leaves open many possibilities for novel designs and new applications. We therefore aim of giving a complete overview on the topic, so as to provide an ordered collection of designs and uses, as well as to spur further development on the subject of the Tamm plasmon for sensing applications.

## Introduction

The so-called Tamm plasmon resonance consists in the enhancement of the surface plasmon resonance (SPR) absorption in a thin metallic layer at the interface with a wavelength-matching bandgap of a photonic crystal (PhC). Specifically, the SPR is a collective oscillation of electrons occurring at the outer surface of a metallic layer, driven by photon-electron coupling, propagating parallel to the surface of a conductive material [1], and appears as an absorption peak around a specific wavelength, which is determined by the plasma frequency of the material. In the case of metallic nanostructures, a localized surface plasmon resonance (LSPR) arises owing to the spatial confinement at a length scale comparable to or smaller than the wavelength of light used to excite the plasmon. Two main properties of SPR have been exploited in sensing for decades [1,2]: the first is the enhancing of the signal in optical measurements, such as in surface-enhanced Raman scattering (SERS) [3] and plasmon-enhanced fluorescence [4,5]. This is due to a strong enhancement of the local electric field that amplifies the probe signal. The other is the change in optical response that stems from either refractive index changes in the light-confining structure or variations in the electron density of the metal. This allows for detecting specific substances and tiny amounts of contaminants, changes in the chemical environment, electromagnetic phenomena, and other localized effects. The use of the Tamm plasmon is a variation on this approach, since in this case the plasmon resonance is buried at the interface between the PhC and the metal. However, if the conductive layer is corrugated and/or nanostructured, the Tamm plasmon provides and exceptional tool for detecting small variations at the outer surface [6]. This has many interesting applications in the field of sensing, and enables the development of a potentially large number of new devices with use in microfluidics, biomedical sciences, microelectronics, and, of course, both integrated and free-standing photonics.

**Photonic Crystals and Plasmonics**

As it has been introduced, the SPR arises from the interaction of an electromagnetic wave with the free electrons of a conductor such as a metal. When this is coupled to a light-confining device, this interaction is enhanced as the radiation intensity is increased close to the interface with the dielectric material. Such a phenomenon can be obtained by pairing a layer of metal, or metal particles, with an interferential mirror, such as a 1-D photonic crystal, also known as a distributed Bragg reflector (DBR). In this case, the coupling of the SPR with the photonic bandgap (PBG), which is the forbidden photon energy band arising from the destructive interference of the periodic DBR layer interfaces having near-unitary reflection, leads to the formation of an optical state known as Tamm plasmon (TP). The name comes from its analogy with the Tamm surface optical state [7], which in turn takes its name by a further analogy to the electronic states at crystal boundaries [8]. Tamm plasmon were first predicted between 2006 by Vinogradov et *al*. [9] and 2007 by Kaliteevski et *al*. [10], and observed experimentally the following year by Sasin et *al* [11]. The Tamm plasmon appears as a drop in the device reflection spectra (and conversely, a peak in their transmission spectra), usually positioned close to the low-energy edge of the photonic crystal PBG (see Fig. 1).

The effect of a dielectric environment on LSPR has also been documented in the context of SERS setups, where single dielectric thin films have been shown to have a substantial effect on the resonance peak of metal layers. For instance, Hong et *al*. [5] investigated the contributions of $TiO_2$ and ITO layers on the absorption spectra of silver films: they showed that the LSPR absorption peak undergoes a marked red-shift as a consequence of the presence of dielectric material, with an increase of the red-shift magnitude for thicker dielectric layers. This effect can be appealing for tuning the spectroscopic properties of metal surfaces, *i.e.* for SERS measurements and similar LSPR-based applications.

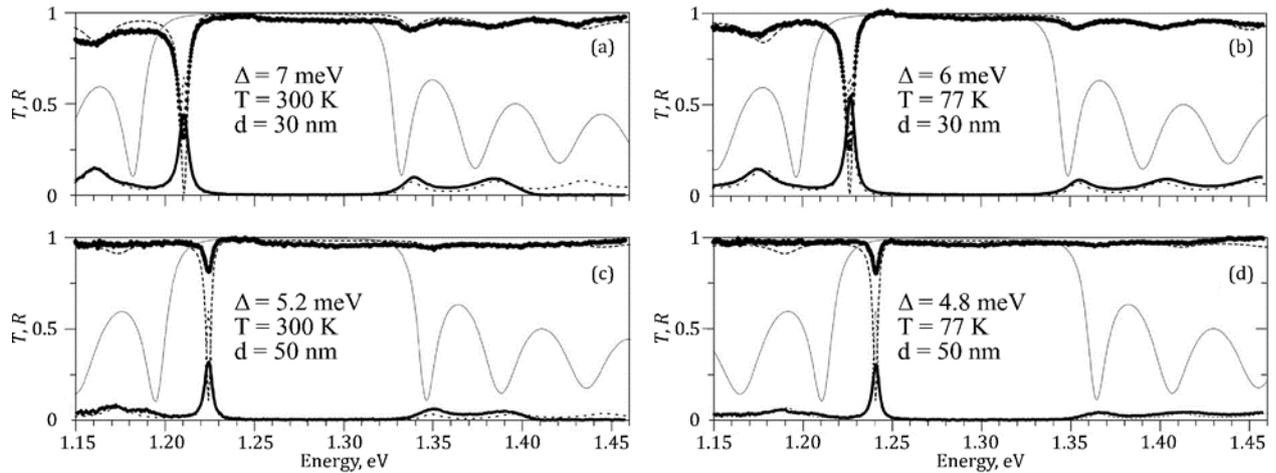

**Figure 1.** First recorded observation of the Tamm plasmon (TP): Transmission and reflection spectra of GaAs/AlAs DBRs covered by layers of gold of thickness [(a) and (b)] d=30 nm and [(c) and (d)] d=50 nm taken at [(a) and (c)] 300 K and [(b) and (d)] 77 K. Circles and solid lines correspond to the measured reflection and transmission spectra, respectively; dashed and dotted lines show the calculated reflection and transmission spectra. Reflection spectra of the DBR uncovered by gold are shown by thin solid lines. Δ is the full width at half maximum of the spectral feature associated with the TP. Reproduced from Sasin et al., Appl. Phys. Lett., 2008, 92, 251112; https://doi.org/10.1063/1.29524866 with the permission of AIP Publishing.

## Design and Fabrication of Tamm Plasmon Devices

TP-based devices play an important role in improving the performance of otherwise purely photonic DBR-based devices employed as refractive index sensors [12]. The conductive layer is critical to the performance of the device, and its thickness can be tuned to maximize the sensitivity of the detector. Das et *al*. have analyzed the response of detectors based on hybrid Tamm plasmon-surface plasmon modes with high-layer count DBRs, and the dependency of their sensitivity on the thickness of the metal layer, in the cases of gold, silver and aluminium [13]: their conclusions showed that layers in the order of tens of nanometers yield the best sensitivities, up to a value over which the Tamm and surface plasmon contributions become uncoupled.

The plasmonic layer, which is as an integral part of the device, requires fine control over its thickness, coverage, and topography. Thermal evaporation is one of the most common techniques for depositing thin layers of metal, however, depending on the layer thickness, the material may present drastically different morphologies due to the formation of nanometer- or micrometer-scale particles. A paper by Gaspar et *al*., in particular, investigated the growth of gold thin films by thermal evaporation and the optical response of the resulting particle layers [14]: as the thickness of the film increases, the particles increase in size as smaller particles join together, and the SPR peak becomes better defined. It is therefore important to note that, in order to have a clear SPR, the particles should exceed sizes of several tens of nanometers, meaning that a gold film for a TP-based device to be performant should have a minimum thickness of 7 nm, according to the results in the aforementioned paper.

In their seminal work [15], Auguié et al. investigated the design optimization of a semiconductor gold-based TP device by determining the critical coupling conditions of light to the TP. Using numerical simulations, they identified the critical coupling conditions for a number of DBR layer pairs equal to 13, with an optimal Au thickness just over 30 nm. The paper provides a thorough mathematical description of the calculations, which is extremely useful to understand the mechanics of light coupling to the TP and therefore to the design of TP-based devices in general, forming a clear basis for the understanding of the working principles of this kind of structures.

Another development was proposed by Symonds et al. in a 2017 paper, wherein the authors show the rise of a high-quality factor "super Tamm plasmon" resonance [16], obtained by manipulating the coupling between the metal layer and the DBR using a dielectric spacer. The article reports on the achievement of a quality factor (defined as the ratio between the position of the resonance peak and its full width at half maximum) over 5000 with the insertion of a 20 nm PMMA spacer (see Fig. 2) and ascribes the reduction of the losses to a reduction of the electric field extension within the metal with respect to the part in the dielectric material. In addition, the authors investigated a structure made of regular Tamm rings that confine the super Tamm structures in their center, which was shown to further improve the quality factor, allowing for the potential further improvement of such devices.

Later models were developed for different designs, such as the recent Ag-coated semiconductor device by Pugh *et al.* realized with a DBR made of undoped GaN and n-doped porous GaN layers [17], or the superlattice device studied by Qiao et al. made out of a silver-coated HfO2/SiO2 with a nanostructured surface [18]. The number of recent works investigating various designs of TP-based devices indicates a high degree of interest in the applications of this technology, as well as the potential versatility of Tamm sensors in a number of different fields.

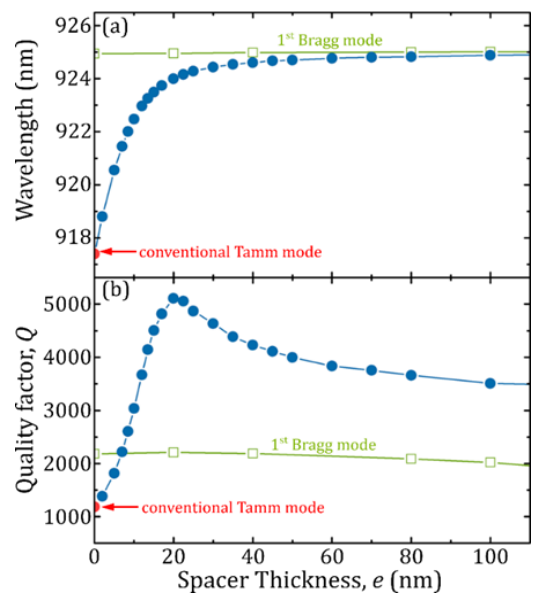

**Figure 2.** Variation of (a) resonance wavelength and (b) quality factor with the spacer thickness calculated by the transfer matrix method. The spacer (n = 1.485) is either inserted between a 65-pair DBR and a 45 nm-thick silver layer (blue circles) or deposited on a 65-pairs DBR without metal (green open squares). Adapted from Symonds et al.[16] (Open access under Creative Commons license CC BY 4.0, https://creativecommons.org/licenses/by/4.0/).

# Applications of Tamm Plasmon detectors

## Porous sensors

The coupling between SPR and PBG in particular has the effect of increasing the sensitivity of the device as compared to simple metal layers and dielectric DBRs, respectively: the TP resonance peak is sensibly narrower than both coupled spectroscopic features by themselves, which makes the shifts more obvious than for broad spectral lineshapes. For instance, mesoporous DBR-based TP devices have been shown to work as effective gas detectors by correlating the TP position to the effective refractive index of the DBR, which is influenced by the refractive index of the surrounding atmosphere, as it changes when gas infiltrates the porous layers [19,20]: while it is possible to detect a shift in the PBG of the DBR alone, the presence of the TP greatly enhances the resonance peak figure of merit, making the measurement more precise and reliable. Similar porous Tamm structures have been investigated in detail in the works of Sansierra et *al*. [19], and – more recently – Juneau-Fecteau et *al*. [21] (Fig. 3).

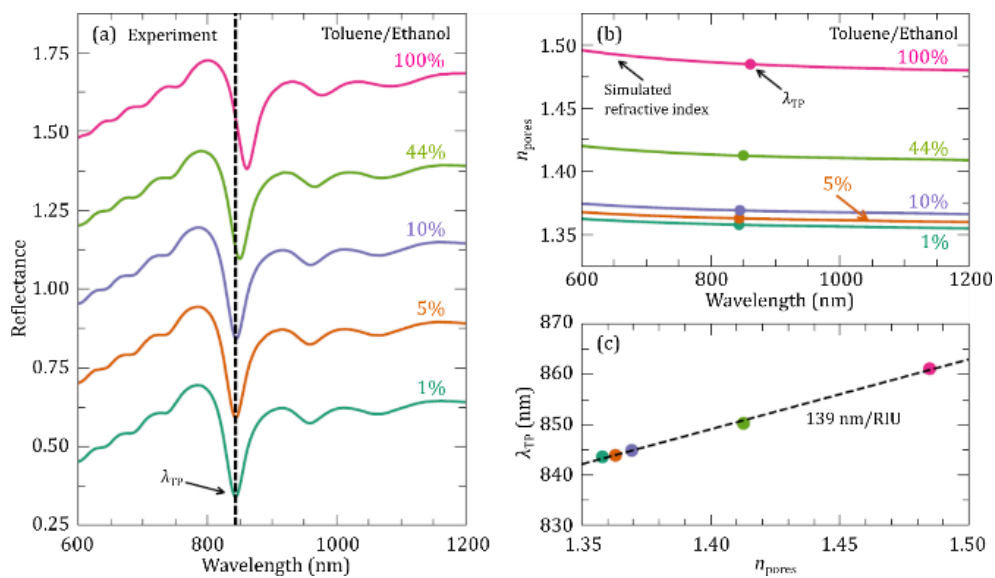

**Figure 3.** (a) Measured reflectance of a porous silicon-based Tamm structure for five volumetric ratios of toluene in ethanol from 1% to 100% of toluene. The curves beyond 1% are shifted by 0.25 in reflectance from each other for clarity. (b) Simulated refractive index for each toluene/ethanol mix according to the Bruggeman effective medium model. (c) Wavelength shift of the TP resonance for each refractive index. Adapted from Juneau-Fecteau et al. [21] (Open Access under Creative Commons license CC BY 4.0, https://creativecommons.org/licenses/by/4.0/).

## Non-porous TP devices

These systems have also been shown to work as detectors for the refractive index of fluids in contact with their surface, as demonstrated by Balevičius et al. in 2020 using total internal reflection techniques [22] (see fig. 4). This is an important result because it proves the possibility of microfluidic applications without the need for porous materials and therefore the limitations of relying on infiltration and the need for pore size control, as well as paving the way for further application of surface-contact measurements.

A 2020 paper by Zaky et al. proposes the model for a gas sensor with very high sensitivity to the refractive index of injected fluids [23]. The device is a cell based on a porous silicon (p-Si) multilayer structure, adding a fixed-thickness spacer for the gas to fill in between the metal layer and the DBR, intended for prism-coupling of the incident light probe: the authors report an expected shift of 40 nm over $2 \cdot 10^{-4}$ refractive index units (RIU), for an optimized device that consists of 8 p-Si layer pairs, a 25 nm silver layer, and a 1 μm spacer, showing a very narrow TP, with full width at half maximum of under 1.4 nm. While these theoretical results represent the best-case scenario, they nevertheless provide interesting insight on design principles to maximize the efficiency and sensitivity of such TP-based detectors, and in particular further applications of spacers for attaining high-quality factors, as previously discussed in Symond et al. [16].

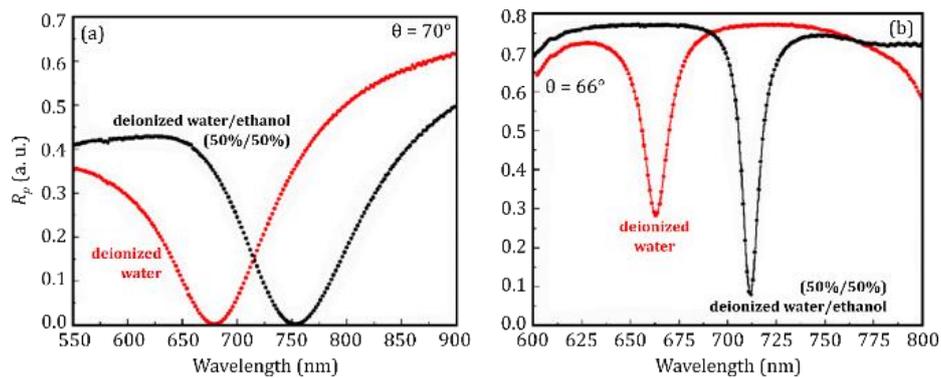

**Figure 4.** Experimental spectra of p-intensity dependence on the wavelength for single SPR (a) and optimized Tamm plasmon (TP) (b) samples before (red curves) and after (black curves) deionized water changed to solution of deionized water/ethanol (50%/50%). Adapted from Balevičius et al. [22] (Open access under Creative Commons license CC BY 4.0, https://creativecommons.org/licenses/by/4.0/).

**Contact devices**

In a similar way to porous sensors, the function of Au-coated non-porous 1-D PhC devices as microfluidic sensors was explored by Shaban et al., where they demonstrated the changes in the TP resonance peak as a function of refractive index [24]. In particular, they used this approach to correlate parameters such as the position and width of the resonance peak with the concentration of glucose within an aqueous solution in contact with the TP-based device. This work is interesting as it provides a solid basis for a potential development of micro- or nanofluidic sensors, without the need for infiltration of the fluid within the DBR layers themselves, which can be a limiting factor in terms of the device response time, and its reusability, and particularly in the case of measurements in liquids.

In further development, TP-based devices have been shown to work as biochemical sensors, in particular in the case of bacterial detection: Paternò et al. have investigated the spectroscopic response of simple DBRs with thin silver layers to the presence of gram-negative bacteria such as *E. coli* [25–27]. In this case, it was shown that the interaction with the silver layer leading to the retention of the bacterial cells on the device surface causes a shift in the PBG (see Fig. 5). It is also assumed that, due to the release of Ag+ ions, the change in free electron density of a silver layer should lead to a shift in the LSPR peak [28], however this resonance is not seen with very thin metal layers (8 nm) and is currently under investigation for thicker ones (20-30 nm), where the TP resonance can be observed clearly.

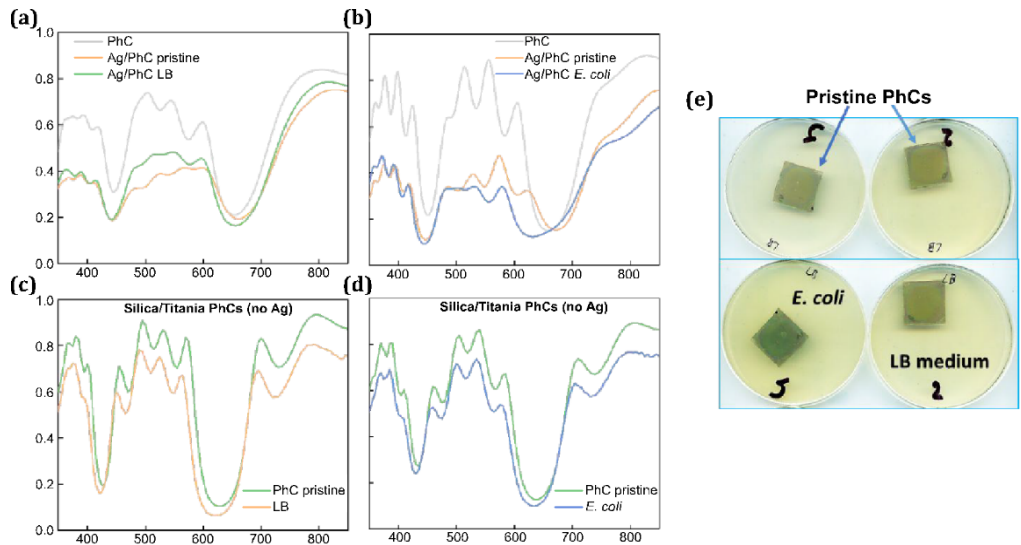

**Figure 5.** (a) Transmittance spectra of 1D PhCs before, after silver deposition and upon exposure to the culture medium and (b) and after exposure to E. coli. (c) Transmittance spectra of the 1D PhCs without silver upon exposure to Luria-Bertani medium and (d) and after exposure to *E. coli*. (e) Colorimetric change of the photonic crystals after contamination with *E. coli*, while exposure to LB does not cause any substantial colorimetric change. Reproduced from Paternò et al.[26] with the authors' permission.

A different design for Tamm devices was investigated by Kumar et *al*. [29]. In such article, the authors showed the fabrication and assessment of a microfluidic TP-based detector made out of two mirrored Tamm multilayer structures, with the analyte flowing within the facing metal layers, effectively forming a spacer between the metal layers. In this case, a double peak appears in place of the standard single-peak reflection drop, as a consequence of the presence of the two TPs, which become coupled, giving rise to both a symmetric hybrid mode and an antisymmetric one. The spectral separation between the reflection minima of these two modes was shown to correlate with the refractive index of the analyte, and this enhanced the sensitivity and accuracy of the entire measurement with respect to a single TP-based structure, matching that of established techniques such as interferometry and SPR sensing, and therefore making such a construction extremely appealing for a wide range of application in micro- and nanofluidics.

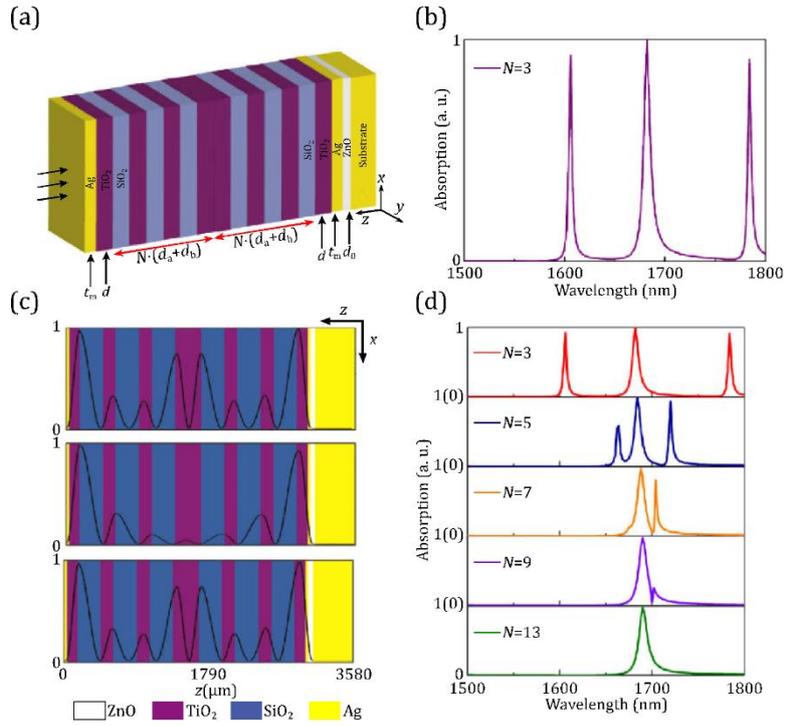

**Figure 6.** Design parameters and performance of a symmetrical dual-Tamm structure. (a) Dual-Tamm structure. (b) Absorption spectrum of incident light wave in near-infrared band. (c) Electric field distribution in the structure at three absorption peaks wavelength. (d) Effect of different DBR period number N on the absorption spectrum in the structure. Adapted from Hu et al.[30] (Open access under the terms of the OSA Open Access Publishing Agreement).

Another dual-Tamm device setup was proposed in the recent work of Hu *et al*. [30], wherein the metallic layers are placed at the opposite ends of specular DBRs, with the dielectric multilayers facing each other. In this case, two layers of the same material and thickness are joined together, effectively forming an optical cavity: the device absorption spectra, therefore, sport three distinct resonance peaks, due to the coupling of the two TPs and the cavity mode, and as the number of DBR layers increases, decoupling the resonances, the three features shift closer together towards the pure TP position (see fig. 6).

A summary of the estimated sensitivities for the TP-based refractive index sensing devices studied in the aforementioned literature is shown in Table 1.

**Table 1.** Experimental and theoretical sensitivities to environmental refractive index changes for various DBR/TP based devices investigated in the compiled scientific literature.

| Reference | Data Type | Design and Materials | Sensitivity (nm/RIU) |
|---|---|---|---|
| Ahmed and Mehaney [20] | Experimental | 25 × (66% p-Si 500 nm / 77% p-Si 1 μm / 85% p-Si 1 μm) , Ag 40 nm<br>25 × (66% p-Si 500 nm / 77% p-Si 1 μm / 85% p-Si 1 μm) , Au 40 nm<br>25 × (66% p-Si 500 nm / 77% p-Si 1 μm / 85% p-Si 1 μm) , Al 40 nm<br>25 × (66% p-Si 500 nm / 77% p-Si 1 μm / 85% p-Si 1 μm) , Pt 40 nm | 5018<br>5092.4<br>5031.5<br>5013 |
| Auguié et al. [15] | Experimental | 5 × (SiO$_2$ 79 nm / TiO$_2$ 104 nm) , Au 23 nm | 55 |
| Balevičius et al. [22] | Experimental | 6 × (TiO$_2$ 120 nm / SiO$_2$ 200 nm) , TiO$_2$ 30 nm, Au 40 nm | 4933 |
| Das et al. [13] | Theoretical | Semi-infinite (TiO$_2$ 75 nm / SiO$_2$ 285 nm) , Au 60 nm<br>Semi-infinite (TiO$_2$ 75 nm / SiO$_2$ 285 nm) , Ag 60 nm<br>Semi-infinite (TiO$_2$ 75 nm / SiO$_2$ 285 nm) , Al 50 nm | ≤ 970<br>≤ 736<br>≤ 738 |
| Hu et al. [30] | Theoretical | Specular [3 × (TiO$_2$ / SiO$_2$) , TiO$_2$ 136 nm , Ag 27 nm] | ≤ 950 |
| Juneau-Fecteau et al. [21] | Experimental | 8 × (37% p-Si 75.5 nm / 59% p-Si 104 nm) , Au 25 nm | 121-139 |
| Kumar et al. [29] | Experimental | Specular [4 × (SiO$_2$ 125 nm / Ta$_2$O$_5$ 60 nm) , Ag 30 nm] , 140 nm cavity (n = 1.33) | ≤ 180 |
| Shaban et al. [24] | Experimental | 10 × (SiN 112 nm / SiO$_2$ 98 nm) , Au 126 nm | ≤ 53.469 |
| Zaky et al. [23] | | 8 × (32% p-Si 200 nm / 74% p-Si 600 nm) , 4 μm spacer , Ag 30 nm | 2600 |

**Electrical sensors**

The TP resonance depends on the electronic characteristics of the metallic layer, therefore metasurfaces can also be employed in conjunction with the DBR and with the use of a liquid crystal to detect changes in electrical tension. Such a device was shown in a 2020 paper by Buchnev et al. [31] , where the plasmonic layer was grown in a periodical nanometric pattern over a Nb2O5-SiO2 DBR surface: submerging the device in liquid crystal and applying a voltage differential through the fluid allowed to tune the TP resonance peak within a 10 nm range in the near-infrared region (see Fig. 7). In particular, the shift occurs for the component of the electric field parallel to the orientation of the liquid crystal molecules. This effect was associated to the change in the local refractive index of the liquid crystal medium near the interface, and therefore of the effective refractive index of the plasmonic layer: this is significant, and the results may pave the way for advances in microelectronics, as such devices may be further improved for the development of voltage detectors, or as a novel instrument to assess the polarization of liquid crystal molecules.

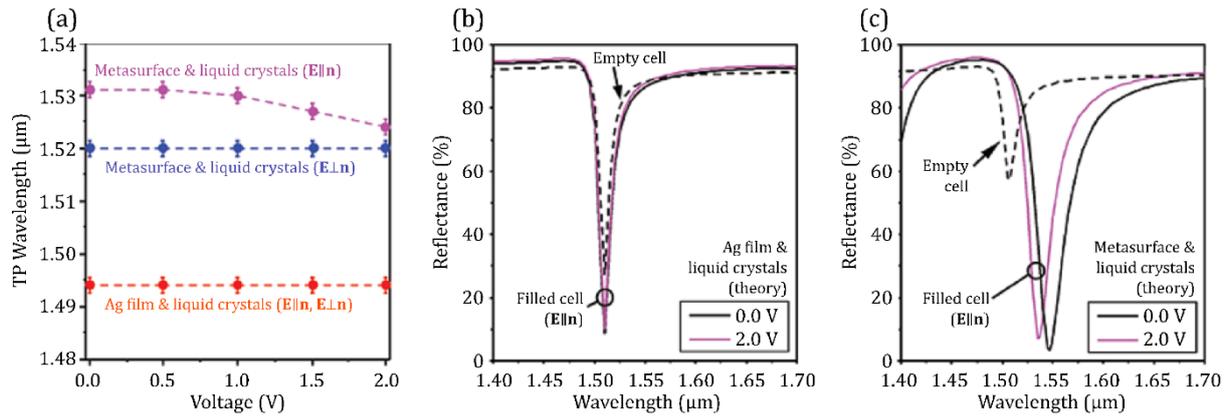

**Figure 7.** Electrical control of Tamm plasmons (experiment and theory). (a) Wavelengths of TP resonance plotted as functions of applied voltage. Dashed curves are a guide for the eye. Panels (b) and (c) display the calculated reflectivity spectra of a continuous silver film and metasurface atop of a DBR integrated with a liquid crystal cell, respectively. Dashed curves show the reflectivity of the structures before the cell was filled with liquid crystal. Solid curves correspond to the planar (black) and homeotropic (magenta) states of the liquid crystal. Adapted from Buchnev et al. [31] (Open access under Creative Commons license CC BY 4.0, https://creativecommons.org/licenses/by/4.0/).

## Conclusions

In the last decades, Tamm plasmon-based devices have been proven to be useful tools for optical sensing applications, with particular use specifically in refractive index sensing and contaminant detection. In this review, we have summarized the main interests for developing TP-based sensors, and the respective designs that make use of this particular phenomenon in different ways. This collection also serves to emphasize the versatility of TP device geometries, and the potential for applications spanning from free-standing, colorimetric detectors, to integrated photonics components for use in micro- and nanofluidics, as well as to give a comprehensive overview of the pending advancements and in TP detector technology. The great array of geometries and materials that can be employed in the design fabrication of TP-based sensors also entails the possibility of combining more of the principles independently explored by the various authors in order to produce more complex devices for an even wider spectrum of implementations.

## CRediT authorship contribution statement

All authors contributed to manuscript drafting and revising, and figure creation.

## Declaration of competing interest

The authors declare that they have no known competing financial interests or personal relationships that could have appeared to influence the work reported in this paper.

## Acknowledgments

This work has been supported by Fondazione Cariplo, grant n° 2018-0979 and n° 2018-0505. F.S. thanks the European Research Council (ERC) under the European Union's Horizon 2020 research and innovation programme (grant agreement No. [816313]).